\documentclass[aps,preprint]{revtex4}%
\usepackage{amsfonts}
\usepackage{amsmath}
\usepackage{amssymb}
\usepackage{graphicx}%
\providecommand{\U}[1]{\protect\rule{.1in}{.1in}}

\begin{document}
\title{Gauge linked time-dependent non-Hermitian Hamiltonians}
\author{F. S. Luiz$^{1}$, M. A. de Ponte$^{2}$ and M. H. Y. Moussa$^{1,3}$}
\affiliation{$^{1}$Instituto de F\'{\i}sica de S\~{a}o Carlos, Universidade de S\~{a}o
Paulo, P.O. Box 369, S\~{a}o Carlos, 13560-970, SP, Brazil}
\affiliation{$^{2}$Universidade Estadual Paulista, Campus Experimental de Itapeva,
18409-010, Itapeva, S\~{a}o Paulo, Brazil}
\affiliation{$^{3}$Centre for Mathematical Science, City University, Northampton Square,
London EC1V 0HB, UK}
\date{}

\begin{abstract}
In this work we address systems described by time-dependent non-Hermitian
Hamiltonians under time-dependent Dyson maps. We shown that when starting from
a given time-dependent non-Hermitian Hamiltonian which is not itself an
observable, an infinite chain of gauge linked time-dependent non-observable
non-Hermitian Hamiltonians can be derived from it. The matrix elements of the
observables associated with all these non observable Hamiltonians are, however,
all linked to each other, and in the particular case where global gauges
exist, these matrix elements becomes all identical to each other. In this
case, therefore, by approaching whatever the Hamiltonian in the chain we can
get information about any other Hamiltonian. We then show that the whole chain
of time-dependent non-Hermitian Hamiltonians collapses to a single
time-dependent non-Hermitian Hamiltonian when, under particular choices for
the time-dependent Dyson maps, the observability of the Hamiltonians is
assured. This collapse thus shows that the observability character of a
non-Hermitian Hamiltonian prevents the construction of the gauge-linked
Hamiltonian chain and, consequently, the possibility of approaching one
Hamiltonian from another.

\end{abstract}

\pacs{32.80.-t, 42.50.Ct, 42.50.Dv}
\maketitle

\section{Introduction}

Non-Hermitian quantum mechanics has receiving increasing attention in the
literature, and since the decisive contributions of \cite{BB} and
\cite{Mostafazadeh}, it has permeated virtually every field of physics
\cite{Grounds}. Experimental observation of $\mathcal{PT}$-symmetry and
$\mathcal{PT}$-symmetry breaking has been reported in a variety of systems
\cite{Zhang}, and a different physical phenomena have been investigated within
$\mathcal{PT}$-symmetric system \cite{Disorder,Localization,Chaos,Solitons}.

Beyond the widely accepted grounds for treating time-independent [or even
time-dependent (TD) \cite{Carla}] non-Hermitian Hamiltonians through
time-independent Dyson maps, recent contributions \cite{Fring,Miled,Luiz} have
advanced the grounds for treating time-independent and specially TD
non-Hermitian Hamiltonians through TD Dyson maps. Although it has been
demonstrated that a TD metric operator can not ensure the unitarity of the
time-evolution simultaneously with the observability of a non-Hermitian
Hamiltonian \cite{Mostafa}, in Ref. \cite{Fring} it has been demonstrated
that, in spite of the non-observability of the Hamiltonian under a TD metric
operator, any other observable associated with this Hamiltonian is derived in
complete analogy with the case where a time-independent Dyson map is
considered. And beyond Ref. \cite{Fring}, in a more recent contribution
\cite{Luiz} a method has been presented which enable us to account for the
unitarity of the time-evolution simultaneously with the observability of a
non-Hermitian Hamiltonian even for a TD Dyson map. The method relies on the
construction of a Schr\"{o}dinger-like equation from which the TD Dyson map is
derived from the TD quasi-Hermitian Hamiltonian itself. Moreover, in spite of
the time-dependence of the Dyson map the method ensures a time-independent
metric operator, a necessary condition for the observability of a
quasi-Hermitian Hamiltonian. Therefore, although in agreement with the main
premise in Refs. \cite{Mostafa,Fring}, that a time-independent metric operator
is needed for assuring the unitarity of the time evolution simultaneously with
the observability of a quasi-Hermitian Hamiltonian, in Ref. \cite{Luiz} a TD
Dyson map is considered, and this is an important point since for a TD
non-Hermitian Hamiltonian, a time-independent Dyson map is a rather
restrictive choice.

In the present contribution we follow the path explored by Refs.
\cite{Fring,Miled,Luiz} to advance some interesting properties derived from
non-Hermitian Hamiltonians under TD metric operators. Working within the main
premise of Ref. \cite{Mostafa,Fring}, that a non-Hermitian Hamiltonian is not
itself an observable when the unitarity of the time evolution is ensured, we
first verify that one can build from this Hamiltonian a whole chain of gauge
linked time-dependent non-observable non-Hermitian Hamiltonians. We
demonstrate that the matrix elements of the observables associated with the
gauge linked non-observable Hamiltonians are all linked to each other, and in
the particular case where global gauges arise, these matrix elements became
all identical to each other. Then, by approaching a given Hamiltonian, we can
obtain information about any other Hamiltonian in the chain and, it is worth
noting that it is immediate to identify that one of the Hamiltonians in the
chain is easier to approach than the other. However, working under the
premises of Ref. \cite{Luiz}, where the Schr\"{o}dinger-like equation is
considered for the derivation of a TD Dyson map, thus enabling us to ensure
the unitarity of the time evolution simultaneously with the observability of a
non-Hermitian Hamiltonian, we automatically prevent the possibility of the
Hamiltonian chain: In other words, the whole chain reduces to a single TD
observable non-Hermitian Hamiltonian, showing that the observability character
prevents the construction of gauge-linked Hamiltonians and observables. In
short, for a TD non-observable non-Hermitian Hamiltonian we can construct a
whole chain of connected\ non-observables non-Hermitian Hamiltonians whose
associated observables are all connected to each other; however, when the
observability of these Hamiltonians are assured, through the construction of
particular Dyson maps from the Schr\"{o}dinger-like equations, the whole chain collapses.

In what follows we first revisit, in Section II, the developments advanced in
Ref. \cite{Fring} to treat TD non-Hermitian Hamiltonians under TD Dyson maps
and metric operators. We then show how to construct from a TD non-Hermitian
Hamiltonian an infinite chain of gauge linked TD non-observable non-Hermitian
Hamiltonians. The observables associated with these non-observables
Hamiltonians are then discussed, specially within the particular case where
global gauge transformations exist. In Section III all the developments in
Section II is revisited now within the construction in Ref. \cite{Luiz} where
a TD non-Hermitian Hamiltonian is itself an observable under a TD Dyson map
(but a time-independent metric operator) derived from the Schr\"{o}dinger-like
equation. Two illustrative examples are given in Section IV, the TD harmonic
oscillator under TD non-Hermitian linear and parametric amplification
processes, and finally, in Section V we present our conclusions.

\section{Time-dependent non-Hermitian systems}

Our starting point is a non-Hermitian TD Hamiltonian $H_{t}\neq H_{t}^{\dag}$
that satisfies the TD Schr\"{o}dinger equation and the TD quasi-Hermiticity
relation \cite{Fring}%
\begin{equation}
H_{t}\psi_{t}=i\hbar\partial_{t}\psi_{t},\text{ \ \ }H_{t}^{\dag}\rho_{t}%
-\rho_{t}H_{t}=i\hbar\partial_{t}\rho_{t},\label{1}%
\end{equation}
respectively. Defining a time-dependent Dyson map $\eta_{t}$ via the relation
$\rho_{t}:=\eta_{t}^{\dag}\eta_{t}$, it follows from (\ref{1}) that the wave
function $\phi_{t}=\eta_{t}\psi_{t}$ satisfies the TD Schr\"{o}dinger equation
for the Hermitian Hamiltonian $h_{t}=h_{t}^{\dag}$ related to $H_{t}$ in a TD
quasi Hermiticity manner{}%
\begin{equation}
h_{t}\phi_{t}=i\hbar\partial_{t}\phi_{t},\text{ \ \ }h_{t}=\eta_{t}H_{t}%
\eta_{t}^{-1}+i\hbar\left(  \partial_{t}\eta_{t}\right)  \eta_{t}%
^{-1}.\label{2}%
\end{equation}
The standard quasi-Hermiticity relations are obtained when $\eta$ and $\rho$
are time-independent.

The time-dependent quasi-Hermiticity relation in Eq. (\ref{1}) ensures that
the time-dependent probabilities in the Hermitian and non-Hermitian systems
are related as%

\begin{equation}
\left\langle \phi_{t}\left\vert \tilde{\phi}_{t}\right.  \right\rangle
=\left\langle \psi_{t}\left\vert \rho_{t}\tilde{\psi}_{t}\right.
\right\rangle :=\left\langle \psi_{t}\left\vert \tilde{\psi}_{t}\right.
\right\rangle _{\rho_{t}}\text{,} \label{3}%
\end{equation}
and consequently, that any observable $o_{t}$ in the Hermitian system has an
observable counterpart
\begin{equation}
O_{t}=\eta_{t}^{-1}o_{t}\eta_{t}\text{,} \label{4}%
\end{equation}
in the non-Hermitian system in complete analogy with the scenario where the
Hamiltonian $H$ and the Dyson map $\eta$ are time-independent operators.

Defining now the two Hilbert spaces $\mathcal{H}(\phi)$ with inner product
$\left\langle \phi_{t}\left\vert \tilde{\phi}_{t}\right.  \right\rangle $ and
$\mathcal{H}(\psi)$ with inner product $\left\langle \psi_{t}\left\vert
\rho_{t}\tilde{\psi}_{t}\right.  \right\rangle =\left\langle \psi
_{t}\left\vert \tilde{\psi}_{t}\right.  \right\rangle _{\rho_{t}}$, it is
easily seen that the operators $u_{t,t^{\prime}}$ and $U_{t,t^{\prime}}=$
$\eta_{t}^{-1}u_{t,t^{\prime}}\eta_{t^{\prime}}$ taking a wave function from
time $t^{\prime}$ to $t$ generate unitarity time-evolution operators, i.e.
preserve probabilities, in those two spaces. We simple verify%
\begin{align}
\left\langle \phi_{t}\left\vert \tilde{\phi}_{t}\right.  \right\rangle  &
=\left\langle u_{t,t^{\prime}}\phi_{t^{\prime}}\right\vert \left.
u_{t,t^{\prime}}\tilde{\phi}_{t^{\prime}}\right\rangle =\left\langle
\phi_{t^{\prime}}\right\vert \left.  u_{t,t^{\prime}}^{\dag}u_{t,t^{\prime}%
}\tilde{\phi}_{t^{\prime}}\right\rangle \nonumber\\
&  =\left\langle \phi_{t^{\prime}}\right\vert \left.  \tilde{\phi}_{t^{\prime
}}\right\rangle ,\label{5}%
\end{align}
and%
\begin{align}
\left\langle \psi_{t}\right\vert \left.  \rho_{t}\tilde{\psi}_{t}%
\right\rangle  &  =\left\langle \eta_{t}^{-1}u_{t,t^{\prime}}\eta_{t^{\prime}%
}\psi_{t^{\prime}}\right\vert \left.  \eta_{t}^{\dag}\eta_{t}\eta_{t}%
^{-1}u_{t,t^{\prime}}\eta_{t^{\prime}}\tilde{\psi}_{t^{\prime}}\right\rangle
\nonumber\\
&  =\left\langle \psi_{t^{\prime}}\right\vert \left.  \eta_{t^{\prime}}^{\dag
}u_{t,t^{\prime}}^{\dag}\left(  \eta_{t}^{\dag}\right)  ^{-1}\eta_{t}^{\dag
}\eta_{t}\eta_{t}^{-1}u_{t,t^{\prime}}\eta_{t^{\prime}}\tilde{\psi}%
_{t}\right\rangle \nonumber\\
&  =\left\langle \psi_{t^{\prime}}\right\vert \left.  \rho_{t^{\prime}}%
\tilde{\psi}_{t^{\prime}}\right\rangle .\label{6}%
\end{align}
The evolution operator $u_{t^{\prime},t}$ satisfies the TD Schr\"{o}dinger
equation (\ref{2}) associated to $h_{t}$ and the standard relations in the
usual manner%
\begin{equation}
h_{t}u_{t,t^{\prime}}=i\hbar\partial_{t}u_{t,t^{\prime}},\text{ \ \ }%
u_{t,t^{\prime}}u_{t^{\prime},t^{\prime\prime}}=u_{t,t^{\prime\prime}},\text{
\ \ and \ \ }u_{t,t}=I,\label{7}%
\end{equation}
and also $U_{t,t^{\prime}}$ is easily shown to satisfy the TD Schr\"{o}dinger
equation (\ref{1}) associated to the Hamiltonian $H_{t}$
\begin{align}
i\hbar\partial_{t}U_{t,t^{\prime}} &  =i\hbar\eta_{t}^{-1}\left(  \partial
_{t}u_{t,t^{\prime}}\right)  \eta_{t^{\prime}}-i\hbar\eta_{t}^{-1}\left(
\partial_{t}\eta_{t}\right)  \eta_{t}^{-1}u_{t,t^{\prime}}\eta_{t^{\prime}%
}\nonumber\\
&  =\eta_{t}^{-1}h_{t}u_{t,t^{\prime}}\eta_{t^{\prime}}-i\hbar\eta_{t}%
^{-1}\left(  \partial_{t}\eta_{t}\right)  U_{t,t^{\prime}}\nonumber\\
&  =\eta_{t}^{-1}h_{t}\eta_{t}U_{t,t^{\prime}}-i\hbar\eta_{t}^{-1}\left(
\partial_{t}\eta_{t}\right)  U_{t,t^{\prime}}\nonumber\\
&  =H_{t}U_{t,t^{\prime}}.\label{Eq8}%
\end{align}
Since observables need to be self-adjoint operators, as in Eq. (\ref{4}), the
Hamiltonian $H_{t}$ satisfying the TD Schr\"{o}dinger equation (\ref{1}) and
generating the time-evolution (\ref{Eq8}) is not an observable quantity as
pointed out in \cite{Mostafazadeh}. Instead the operator
\begin{equation}
H_{t}^{\prime}=\eta_{t}^{-1}h_{t}\eta_{t}=H_{t}+i\hbar\eta_{t}^{-1}%
\partial_{t}\eta_{t},\label{9}%
\end{equation}
is an observable quantity in the Hilbert space $\mathcal{H}(\psi)$. The
operator $H_{t}^{\prime}$ is of course not a Hamiltonian in that space, but it
can be used to set up a system of new TD quasi-Hermitian operators%
\begin{subequations}
\begin{align}
H_{t}^{\prime}\psi_{t}^{\prime} &  =i\hbar\partial_{t}\psi_{t}^{\prime},\text{
\ \ }\left(  H_{t}^{\prime}\right)  ^{\dag}\rho_{t}^{\prime}-\rho_{t}^{\prime
}H_{t}^{\prime}=i\hbar\partial_{t}\rho_{t}^{\prime}\text{,}\label{10a}\\
h_{t}^{\prime}\phi_{t}^{\prime} &  =i\hbar\partial_{t}\phi_{t}^{\prime},\text{
\ \ }h_{t}^{\prime}=\eta_{t}^{\prime}H_{t}^{\prime}\left(  \eta_{t}^{\prime
}\right)  ^{-1}+i\hbar\left(  \partial_{t}\eta_{t}^{\prime}\right)  \left(
\eta_{t}^{\prime}\right)  ^{-1}\text{,}\label{10b}%
\end{align}
with $\phi_{t}^{\prime}=\eta_{t}^{\prime}\psi_{t}^{\prime},$ $\rho_{t}%
^{\prime}:=\left(  \eta_{t}^{\prime}\right)  ^{\dag}\eta_{t}^{\prime}$ and new
Hilbert spaces $\mathcal{H}(\phi^{\prime})$ and $\mathcal{H}(\psi^{\prime})$.

\subsection{Gauge symmetrically linked Hamiltonian chain}

Apart from being linked by $H_{t}^{\prime}$, at this point the two systems
$\mathcal{H}(\phi),\mathcal{H}(\psi)$ and $\mathcal{H}(\phi^{\prime
}),\mathcal{H}(\psi^{\prime})$ are unrelated. In order to achieve that we
assume that $\mathcal{H}(\phi)$ and $\mathcal{H}(\phi^{\prime})$ are related
to each other by a gauge transformation. In other words we assume that
$\phi_{t}^{\prime}$ and $\phi_{t}$ are related to each other by a unitary
operator $A_{t}$ as $\phi_{t}^{\prime}$ $=A_{t}\phi_{t}$. The substitution of
this relation into (\ref{10b}) leads to the standard expression
\end{subequations}
\begin{equation}
h_{t}^{\prime}=A_{t}h_{t}A_{t}^{-1}+i\hbar\left(  \partial_{t}A_{t}\right)
A_{t}^{-1}, \label{12}%
\end{equation}
and consequently to
\begin{equation}
i\hbar\partial_{t}A_{t}=h_{t}^{\prime}A_{t}-A_{t}h_{t}, \label{12a}%
\end{equation}
where $h_{t}^{\prime}$ follows by substituting Eq. (\ref{9}) into Eq.
(\ref{10b}), thus giving
\begin{equation}
h_{t}^{\prime}=\eta_{t}^{\prime}\eta_{t}^{-1}h_{t}\eta_{t}\left(  \eta
_{t}^{\prime}\right)  ^{-1}+i\hbar\left(  \partial_{t}\eta_{t}^{\prime
}\right)  \left(  \eta_{t}^{\prime}\right)  ^{-1}. \label{12b}%
\end{equation}
Thus we have now related the wavefunction of all four Hilbert spaces to each
other%
\begin{equation}
\psi_{t}^{\prime}=\left(  \eta_{t}^{\prime}\right)  ^{-1}\phi_{t}^{\prime
}=\left(  \eta_{t}^{\prime}\right)  ^{-1}A_{t}\phi_{t}=\left(  \eta
_{t}^{\prime}\right)  ^{-1}A_{t}\eta_{t}\psi_{t}. \label{13}%
\end{equation}
Similarly we may now also relate all four unitary time-evolution operators to
each other as
\begin{align}
U_{t,t^{\prime}}^{\prime}  &  =\left(  \eta_{t}^{\prime}\right)
^{-1}u_{t,t^{\prime}}^{\prime}\eta_{t^{\prime}}^{\prime}\nonumber\\
&  =\left(  \eta_{t}^{\prime}\right)  ^{-1}A_{t}u_{t,t^{\prime}}A_{t^{\prime}%
}^{-1}\eta_{t^{\prime}}^{\prime}\nonumber\\
&  =\left(  \eta_{t}^{\prime}\right)  ^{-1}A_{t}\eta_{t}U_{t,t^{\prime}}%
\eta_{t^{\prime}}^{-1}A_{t^{\prime}}^{-1}\eta_{t^{\prime}}^{\prime}.
\label{14}%
\end{align}

Back to Eq. (\ref{10b}) we note that $H_{t}^{\prime}$ is not an observable in
the space $\mathcal{H}(\psi^{\prime})$, contrarily to the operator
\begin{equation}
H_{t}^{\prime\prime}=\left(  \eta_{t}^{\prime}\right)  ^{-1}h_{t}^{\prime}%
\eta_{t}^{\prime}=H_{t}^{\prime}+i\hbar\left(  \eta_{t}^{\prime}\right)
^{-1}\partial_{t}\eta_{t}^{\prime},\label{14a}%
\end{equation}
which can then be used to set up another pair of Schr\"{o}dinger equations
\begin{subequations}
\label{15}%
\begin{align}
H_{t}^{\prime\prime}\psi_{t}^{\prime\prime} &  =i\hbar\partial_{t}\psi
_{t}^{\prime\prime},\text{ \ \ }\left(  H_{t}^{\prime\prime}\right)  ^{\dag
}\rho_{t}^{\prime\prime}-\rho_{t}^{\prime\prime}H_{t}^{\prime\prime}%
=i\hbar\partial_{t}\rho_{t}^{\prime\prime}\text{,}\label{15a}\\
h_{t}^{\prime\prime}\phi_{t}^{\prime\prime} &  =i\hbar\partial_{t}\phi
_{t}^{\prime\prime},\text{ \ \ }h_{t}^{\prime\prime}=\eta_{t}^{\prime\prime
}H_{t}^{\prime\prime}\left(  \eta_{t}^{\prime\prime}\right)  ^{-1}+i\left(
\partial_{t}\eta_{t}^{\prime\prime}\right)  \left(  \eta_{t}^{\prime\prime
}\right)  ^{-1}\text{,}\label{15b}%
\end{align}
thus defining another pair of Hilbert spaces $\mathcal{H}(\phi^{\prime\prime
}),\mathcal{H}(\psi^{\prime\prime})$. To link together the two systems
$\mathcal{H}(\phi^{\prime}),\mathcal{H}(\psi^{\prime})$ and $\mathcal{H}%
(\phi^{\prime\prime}),\mathcal{H}(\psi^{\prime\prime})$, we assume that
$\phi_{t}^{\prime\prime}$ $=A_{t}^{\prime}\phi_{t}^{\prime}$, with
$A_{t}^{\prime}$ being another unitary operator, such that, similarly to Eqs.
(\ref{12}), (\ref{12a}), and (\ref{12b}), we now have
\end{subequations}
\begin{subequations}
\begin{align}
h_{t}^{\prime\prime} &  =A_{t}^{\prime}h_{t}^{\prime}\left(  A_{t}^{\prime
}\right)  ^{-1}+i\hbar\left(  \partial_{t}A_{t}^{\prime}\right)  \left(
A_{t}^{\prime}\right)  ^{-1}\label{16a}\\
&  =\eta_{t}^{\prime\prime}\left(  \eta_{t}^{\prime}\right)  ^{-1}%
h_{t}^{\prime}\eta_{t}^{\prime}\left(  \eta_{t}^{\prime\prime}\right)
^{-1}+i\hbar\left(  \partial_{t}\eta_{t}^{\prime\prime}\right)  \left(
\eta_{t}^{\prime\prime}\right)  ^{-1},\label{16b}%
\end{align}
and consequently
\end{subequations}
\begin{equation}
i\hbar\partial_{t}A_{t}^{\prime}=h_{t}^{\prime\prime}A_{t}^{\prime}%
-A_{t}^{\prime}h_{t}^{\prime}.\label{16c}%
\end{equation}

We can then build a whole chain of Hamiltonians starting from $H_{t}$ and
going through $H_{t}^{\prime},H_{t}^{\prime\prime},...$ with their associated
Hermitian counterparts $h_{t},h_{t}^{\prime},h_{t}^{\prime\prime},...$ derived
through the time-dependent Dyson maps $\eta_{t},\eta_{t}^{\prime},\eta
_{t}^{\prime\prime},...$ and the wave functions $\phi_{t},\phi_{t}^{\prime
},\phi_{t}^{\prime\prime},...$ related to each other by the unitary operators
$A_{t},A_{t}^{\prime},...$. This construction, leading to the Hamiltonians
derived from $H_{t}$: $H_{t}\rightarrow H_{t}^{\prime}=H_{t}+i\hbar\eta
_{t}^{-1}\partial_{t}\eta_{t}\rightarrow H_{t}^{\prime\prime}=H_{t}^{\prime
}+i\hbar\left(  \eta_{t}^{\prime}\right)  ^{-1}\partial_{t}\eta_{t}^{\prime
}\rightarrow...$, can also be carried out in reverse by looking for the
derivation of $H_{t}$ itself from\thinspace%
\begin{equation}
\overset{\_}{H}_{t}=H_{t}-i\hbar\left(  \overset{\_}{\eta_{t}}\right)
^{-1}\partial_{t}\overset{\_}{\eta}_{t},\label{eq17a}%
\end{equation}
and then the derivation of $\overset{\_}{H}_{t}$ from
\begin{equation}
\overset{\_}{\overset{\_}{H}}_{t}=\overset{\_}{H}_{t}-i\hbar\left(
\overset{\_}{\overset{\_}{\eta}}_{t}\right)  ^{-1}\partial_{t}\overset
{\_}{\overset{\_}{\eta}}_{t}\text{,}\label{eq17b}%
\end{equation}
and so on, thus leading to the infinite chain of non-Hermitian Hamiltonians
\begin{equation}
...\rightarrow\overset{\_}{\overset{\_}{H}}_{t}\rightarrow\overset{\_}{H}%
_{t}\rightarrow H_{t}\rightarrow H_{t}^{\prime}\rightarrow H_{t}^{\prime
\prime}\rightarrow...\text{,}\label{eq18}%
\end{equation}
which are related with their Hermitian counterparts
\begin{equation}
...\rightarrow\bar{h}_{t}=\overset{\_}{\eta}_{t}H_{t}\left(  \overset{\_}%
{\eta}_{t}\right)  ^{-1}\rightarrow h_{t}=\eta_{t}H_{t}^{\prime}\eta_{t}%
^{-1}\rightarrow h_{t}^{\prime}=\eta_{t}^{\prime}H_{t}^{\prime\prime}\left(
\eta_{t}^{\prime}\right)  ^{-1}\rightarrow...\text{,}\label{19}%
\end{equation}
where, as to be discussed bellow, the Hamiltonian $\bar{h}_{t}$, on the border
between the prime and the bar Hamiltonians, differs from all others because it
does not involve a time derivative of the Dyson map operator.

Thus, following the same procedure leading from $H_{t}$ to $H_{t}^{\prime}$
through $\eta_{t}$, and so on, by defining $\overset{\_}{H}_{t}$ in the way
written above we immediately obtain $H_{t}$ through $\overset{\_}{\eta}_{t}$
and thus all the Hamiltonians preceding $\overset{\_}{H}_{t}$ as given by the
chain in Eq. (\ref{eq18}). The Hamiltonian $\overset{\_}{H}_{t}$ is then
constructed from $H_{t}$ and the Dyson map $\left(  \overset{\_}{\eta}%
_{t}\right)  ^{-1}$ \emph{previously} to the transformation that this map
performs on the Schr\"{o}dinger equation for $\overset{\_}{H}_{t}$ leading to
that for $H_{t}$, the same applying for all the Hamiltonians preceding $H_{t}%
$. Differently, the Hamiltonian $H_{t}^{\prime}$ is constructed from $H_{t}$
and the Dyson map $\eta_{t}$ \emph{afterwards} the transformation this map
performs on the Schr\"{o}dinger equation for $H_{t}$, the same applying for
all the Hamiltonians following from $H_{t}$. However, for both cases, the
Hamiltonians preceding $H_{t}$ or following from $H_{t}$, they are fully
determined only after we have computed the time-dependent parameters defining
their respective Dyson maps through the Hermiticity of the required Hermitian counterparts.

\subsection{Observables}

Lets us now turn to the observables $...,\bar{O}_{t},O_{t},O_{t}^{\prime},...$
associated with the non-Hermitian Hamiltonians $...,\bar{H}_{t},H_{t}%
,H_{t}^{\prime},...$ composing the chain in Eq. (\ref{eq18}). Considering, for
example, the observables associated with $H_{t}$, given by Eq. (\ref{4}),
their matrix elements in the space $\mathcal{H}(\psi)$ are related to the
matrix elements of their Hermitian counterparts $o_{t}$ in the space
$\mathcal{H}(\phi)$, as well as in that preceding it, $\mathcal{H}(\bar{\phi
})$, or following it, $\mathcal{H}(\phi^{\prime})$, through the gauge
operators $\bar{A}_{t}$ and $A_{t}$ in the form%
\begin{align}
\left\langle \psi_{t}\left\vert O_{t}\tilde{\psi}_{t}\right.  \right\rangle
_{\rho_{t}} &  =\left\langle \phi_{t}\left\vert o_{t}\tilde{\phi}_{t}\right.
\right\rangle =\left\langle \bar{\phi}_{t}\left\vert \left(  \bar{A}%
_{t}\right)  ^{\dag}o_{t}\bar{A}_{t}\overset{\mathbf{\symbol{126}}}{\bar{\phi
}}_{t}\right.  \right\rangle \nonumber\\
&  =\left\langle \phi_{t}^{\prime}\left\vert A_{t}o_{t}A_{t}^{\dag}\tilde
{\phi}_{t}^{\prime}\right.  \right\rangle .\label{18}%
\end{align}
Similar relations hold for the matrix elements of all other observables
$...,\bar{O}_{t},O_{t}^{\prime},...$, related to the non-Hermitian
Hamiltonians $...,\bar{H}_{t},H_{t}^{\prime},...$. In the same way that the
matrix elements of the observables $...,\bar{O}_{t},O_{t},O_{t}^{\prime},...$
can be computed in whatever the Hilbert space $...,\mathcal{H}(\bar{\phi
}),\mathcal{H}(\phi),\mathcal{H}(\phi^{\prime}),...$, these matrix elements
are all connected to each other, since the space states as well as the
observables are also all connected to each other [the former as given, for
example, by Eq. (\ref{13}), and the latter as given by $O_{t}^{\prime}=\left(
\eta_{t}^{\prime}\right)  ^{-1}\eta_{t}O_{t}\eta_{t}^{-1}\eta_{t}^{\prime}$].
This shows that by approaching a given Hamiltonian in the chain, we thus
obtain information about any other Hamiltonian, and considering again the
observable $O_{t}$, it is immediate to relate its matrix elements in its own
space $\mathcal{H}(\psi)$ with those computed for example in $\mathcal{H}%
(\psi^{\prime})$ as%
\begin{align}
\left\langle \psi_{t}\left\vert O_{t}\tilde{\psi}_{t}\right.  \right\rangle
_{\rho_{t}} &  =\left\langle \psi_{t}^{\prime}\left\vert \left(  \eta
_{t}^{\prime}\right)  ^{\dag}A_{t}\eta_{t}O_{t}\eta_{t}^{-1}A_{t}^{\dag}%
\eta_{t}^{\prime}\tilde{\psi}_{t}^{\prime}\right.  \right\rangle \nonumber\\
&  =\left\langle \psi_{t}^{\prime}\left\vert \left(  \eta_{t}^{\prime}\right)
^{\dag}A_{t}\eta_{t}^{\prime}O_{t}^{\prime}\left(  \eta_{t}^{\prime}\right)
^{-1}A_{t}^{\dag}\eta_{t}^{\prime}\tilde{\psi}_{t}^{\prime}\right.
\right\rangle .\label{Eq18}%
\end{align}

Regarding the matrix elements of the Hamiltonian $H_{t}^{\prime}$ in the space
$\mathcal{H}(\psi)$, where it is an observable, they are related to the matrix
elements of $h_{t}$ in the space $\mathcal{H}(\phi)$, as well as in that
preceding it, $\mathcal{H}(\bar{\phi})$, or following, $\mathcal{H}%
(\phi^{\prime})$, through the gauge operators $\bar{A}_{t}$ and $A_{t}%
^{\prime}$:
\begin{align}
\left\langle \psi_{t}\left\vert H_{t}^{\prime}\tilde{\psi}_{t}\right.
\right\rangle _{\rho_{t}} &  =\left\langle \phi_{t}\left\vert h_{t}\tilde
{\phi}_{t}\right.  \right\rangle =\left\langle \bar{\phi}_{t}\left\vert
\left(  \bar{A}_{t}\right)  ^{\dag}h_{t}\bar{A}_{t}\overset
{\mathbf{\symbol{126}}}{\bar{\phi}}_{t}\right.  \right\rangle \nonumber\\
&  =\left\langle \phi_{t}^{\prime}\left\vert \left(  A_{t}^{\dag}\right)
^{-1}h_{t}\left(  A_{t}\right)  ^{-1}\tilde{\phi}_{t}^{\prime}\right.
\right\rangle \text{.}\label{Eq19}%
\end{align}
The matrix elements of the Hamiltonian $H_{t}^{\prime}$ in the space
$\mathcal{H}(\psi^{\prime})$ are related to the matrix elements of
$h_{t}^{\prime}$ and $h_{t}$ in the space $\mathcal{H}(\phi^{\prime})$ in the
form%
\begin{align}
\left\langle \psi_{t}^{\prime}\right\vert \left.  H_{t}^{\prime}\tilde{\psi
}_{t}^{\prime}\right\rangle _{\rho_{t}^{\prime}} &  =\left\langle \phi
_{t}^{\prime}\right\vert \left.  \eta_{t}^{\prime}\left[  \left(  \eta
_{t}^{\prime}\right)  ^{-1}h_{t}^{\prime}\eta_{t}^{\prime}-i\hbar\left(
\eta_{t}^{\prime}\right)  ^{-1}\partial_{t}\eta_{t}^{\prime}\right]  \left(
\eta_{t}^{\prime}\right)  ^{-1}\tilde{\phi}_{t}^{\prime}\right\rangle
\nonumber\\
&  =\left\langle \phi_{t}^{\prime}\right\vert \left.  \eta_{t}^{\prime}%
\eta_{t}^{-1}h_{t}\eta_{t}\left(  \eta_{t}^{\prime}\right)  ^{-1}\tilde{\phi
}_{t}^{\prime}\right\rangle ,\label{Eq20}%
\end{align}
clearly showing that\textbf{ }$H_{t}^{\prime}$ is not an observable in its own
space\textbf{ }$H(\psi^{\prime})$\textbf{.}

\subsection{Global gauge transformations}

From Eqs. (\ref{12a})\ and (\ref{16c}) we conclude that under global gauge
transformations, where all the operators $...,\bar{A}_{t},A_{t},A_{t}^{\prime
},...$ are time-dependent or constant phase factors, proportional to the
identity, such that $i\hbar\partial_{t}A_{t}=\left[  h_{t}^{\prime}%
-h_{t}\right]  A_{t}$, it follows that%
\begin{equation}
A_{t}=A_{t^{\prime}}\exp\left(  -\frac{i}{\hbar}%
{\textstyle\int\nolimits_{t^{\prime}}^{t}}
\left(  h_{\tau}^{\prime}-h_{\tau}\right)  d\tau\right)  ,\label{Eq21}%
\end{equation}
with similar expressions for $A_{t}^{\prime},A_{t}^{\prime\prime},...$.
Therefore, global gauge operators $A_{t}^{\prime},A_{t}^{\prime\prime},...,$
demand the neighboring Hermitian Hamiltonians to differ from each other only
by a $\mathcal{C}$-number, i.e., $h_{t}^{\prime}-h_{t}=\mathcal{C}_{t}$,
$h_{t}^{\prime\prime}-h_{t}^{\prime}=\mathcal{C}_{t}^{\prime}$,..., such that
$A_{t}=A_{t^{\prime}}\exp\left[  -\frac{i}{\hbar}%
{\textstyle\int\nolimits_{t^{\prime}}^{t}}
\mathcal{C}_{\tau}d\tau\right]  $, and so on. Moreover, global gauges also
demand the Dyson maps to satisfy equations of the form%
\begin{equation}
\partial_{t}\eta_{t}^{\prime}=-\frac{i}{\hbar}\eta_{t}^{\prime}\left[  \left(
\eta_{t}^{\prime}\right)  ^{-1}h_{t}\eta_{t}^{\prime}-\eta_{t}^{-1}h_{t}%
\eta_{t}+\mathcal{C}_{t}\right]  \text{,}\label{Eq22}%
\end{equation}
with similar expressions for all other Dyson maps.

We thus verify that, under global gauge transformations the matrix elements of
the observable $O_{t}$ in the space $\mathcal{H}(\psi)$, as given by Eq.
(\ref{18}), are related to the matrix elements of their Hermitian counterparts
$o_{t}$ in the spaces $\mathcal{H}(\phi)$, $\mathcal{H}(\bar{\phi})$, and
$\mathcal{H}(\phi^{\prime})$, in the simplified form
\begin{align}
\left\langle \psi_{t}\left\vert O_{t}\tilde{\psi}_{t}\right.  \right\rangle
_{\rho_{t}}  &  =\left\langle \phi_{t}\left\vert o_{t}\tilde{\phi}_{t}\right.
\right\rangle =\left\langle \bar{\phi}_{t}\left\vert o_{t}\overset
{\mathbf{\symbol{126}}}{\bar{\phi}}_{t}\right.  \right\rangle \nonumber\\
&  =\left\langle \phi_{t}^{\prime}\left\vert o_{t}\tilde{\phi}_{t}^{\prime
}\right.  \right\rangle . \label{Eq23}%
\end{align}
In addition, the matrix elements of all the observables $...,\bar{O}_{t}%
,O_{t},O_{t}^{\prime},...$(associated with the non-Hermitian Hamiltonians
$...,\bar{H}_{t},H_{t},H_{t}^{\prime},...$), in their respective spaces
$...\mathcal{H}(\bar{\psi}),\mathcal{H}(\psi),\mathcal{H}(\psi^{\prime}),...$,
all equal each other under global gauge transformations:%
\begin{align}
\left\langle \bar{\psi}_{t}\left\vert \bar{O}_{t}\overset{\mathbf{\symbol{126}%
}}{\bar{\psi}}_{t}\right.  \right\rangle _{\bar{\rho}_{t}}  &  =\left\langle
\psi_{t}\right\vert \left.  O_{t}\tilde{\psi}_{t}\right\rangle _{\rho_{t}%
}=\left\langle \psi_{t}^{\prime}\right\vert \left.  O_{t}^{\prime}\tilde{\psi
}_{t}^{\prime}\right\rangle _{\rho_{t}^{\prime}}\nonumber\\
&  =\left\langle \phi_{t}\left\vert o_{t}\tilde{\phi}_{t}\right.
\right\rangle \text{.} \label{Eq24}%
\end{align}

Regarding the matrix elements of the Hamiltonian $H_{t}^{\prime}$ in the the
space $\mathcal{H}(\psi)$, under global gauges they are related to the matrix
elements of $h_{t}$ in the space $\mathcal{H}(\phi)$, $\mathcal{H}(\bar{\phi
})$, and $\mathcal{H}(\phi^{\prime})$, in the form
\begin{align}
\left\langle \psi_{t}\left\vert H_{t}^{\prime}\tilde{\psi}_{t}\right.
\right\rangle _{\rho_{t}} &  =\left\langle \phi_{t}\left\vert h_{t}\tilde
{\phi}_{t}\right.  \right\rangle =\left\langle \bar{\phi}_{t}\left\vert
h_{t}\overset{\mathbf{\symbol{126}}}{\bar{\phi}}_{t}\right.  \right\rangle
\nonumber\\
&  =\left\langle \phi_{t}^{\prime}\left\vert h_{t}\tilde{\phi}_{t}^{\prime
}\right.  \right\rangle \text{.}\label{Eq25}%
\end{align}
whereas the matrix elements of the Hamiltonian $H_{t}^{\prime}$ in the space
$\mathcal{H}(\psi^{\prime})$ are still given by Eq. (\ref{Eq20}).

\subsection{Practical Application of a Gauge Linked Hamiltonian Chain}

Before addressing the method of constructing observables TD non-Hermitian
Hamiltonians which simultaneously imply the unitarity of the Schr\"{o}dinger
time-evolution \cite{Luiz}, it is worth stressing that the Hamiltonian chain
has a very clear practical application: As long as the matrix elements of
their associated observables are all linked together, it becomes simpler to
pick up the Hamiltonian $\bar{H}_{t}$ among all those in the chain. In fact,
the computation of its Hermitian counterpart $\bar{h}_{t}=\overset{\_}{\eta
}_{t}H_{t}\left(  \overset{\_}{\eta}_{t}\right)  ^{-1}$, unlike all other
Hermitian counterparts, does not involve a time derivative of the
corresponding Dyson map, what is generally a difficult task demanding the
Gauss decomposition of this TD operator. In the two illustrative examples
given bellow ---the TD harmonic oscillator under TD linear and nonlinear
amplification processes--- we explore this practical feature.

\section{Observability of the non-Hermitian Hamiltonians in their own spaces
simultaneously to the unitarity of the time evolution}

It has been demonstrated in Ref. \cite{Mostafazadeh} that a TD Dyson map can
not ensure the unitarity of the time-evolution simultaneously with the
observability of the Hamiltonian. The developments in Ref. \cite{Mostafazadeh}
has been extended to demonstrate that despite the nonobservabiltiy of the
non-Hermitian Hamiltonian under a TD Dyson map, a TD Dyson equation and a TD
quasi-Hermiticity relation can be solved consistently, showing that any other
observable in the non-Hermitian system is derived in complete analogy with the
time-independent scenario \cite{Fring}. Solutions to the proposed TD Dyson
equation and quasi-Hermiticity relation have been presented in the literature
\cite{Fring,Miled}.

More recently, in Ref. \cite{Luiz} an strategy has been presented for the
derivation of a time-dependent Dyson map which ensures simultaneously the
unitarity of the time evolution and the observability of a quasi-Hermitian
Hamiltonian. This time-dependent Dyson map is derived deterministicaly, except
for its initial condition, through a Schr\"{o}dinger-like equation governed by
the non-Hermitian Hamiltonian itself. The Schr\"{o}dinger-like equation
follows by imposing, in the Eq. (\ref{2}) for $h_{t}$, the gauge-like term
$i\left(  \partial_{t}\eta_{t}\right)  \eta_{t}^{-1}$ to be equal to $\eta
_{t}H_{t}\eta_{t}^{-1}$, thus leading to%
\begin{equation}
i\hbar\partial_{t}\eta_{t}=\eta_{t}H_{t}.\label{eq5}%
\end{equation}
which is indeed similar to the Schr\"{o}dinger equation written in the dual
Hilbert space. Evidently, the above constructed equation ensures the
similarity transformation%
\begin{equation}
h_{t}=2\eta_{t}H_{t}\eta_{t}^{-1},\label{eq}%
\end{equation}
and by demanding $h_{t}$ to be Hermitian, we derive the quasi-Hermiticity
relation%
\begin{equation}
H_{t}^{\dag}\rho_{t}=\rho_{t}H_{t},\label{Eq5}%
\end{equation}
which consistently implies the time-independency of the metric operator
$\rho_{t}=\rho(t_{0})$ in spite of the time-dependency of the Dyson map. In
fact, the time-independency of the metric is a necessary condition for the
observability of the non-Hermitian $H_{t}$.

Here, considering our gauge linked quasi-Hermitian Hamiltonian chain, when
imposing the Schr\"{o}dinger-like equation of the form (\ref{eq5}) for all the
TD Dyson maps (i.e., $i\hbar\partial_{t}\eta_{t}^{\prime}=\eta_{t}^{\prime
}H_{t}^{\prime}$, $i\hbar\partial_{t}\eta_{t}^{\prime\prime}=\eta_{t}%
^{\prime\prime}H_{t}^{\prime\prime}$, ...), we verify that all the
non-Hermitian Hamiltonian in the chain (\ref{eq18}) automatically reduces to a
single Hamiltonian $H_{t}$ apart from constant factors; to be precise, we
obtain $...,\overset{\_}{\overset{\_}{H}}_{t}=H_{t}/4$, $\overset{\_}{H}%
_{t}=H_{t}/2$, $H_{t}^{\prime}=2H_{t}$, $H_{t}^{\prime\prime}=4H_{t}$, $...$ .
We have thus found here the interesting property that an observable
quasi-Hermitian Hamiltonian is unique, whereas the lack of its observability
enable us to construct the whole chain of gauge symmetrically linked
quasi-Hermitian Hamiltonians $...,\overset{\_}{\bar{H}}_{t},H_{t}%
,H_{t}^{\prime},...$, with the associated observables$...,\bar{O}_{t}%
,O_{t},O_{t}^{\prime},...$. In the case where global gauge transformations are
required, the matrix elements of all the observables in their respective
spaces equal each other, as in Eq. (\ref{24}), despite the fact that the
Hamiltonians themselves are not observables.

\section{The linear time-dependent non-Hermitian Hamiltonian for bosonic
operators}

As an illustrative example of the theory presented above we consider a TD
harmonic oscillator under a non-Hermitian linear amplification process,
described by the Hamiltonian ($\hbar=1$)%

\begin{equation}
H_{t}=\omega_{t}a^{\dagger}a+\alpha_{t}a+\beta_{t}a^{\dagger}\text{,}
\label{26}%
\end{equation}
where $a$ and $a^{\dagger}$ are bosonic annihilation and creation operators
and the TD parameters $\omega_{t},\alpha_{t},\beta_{t}$ are complex functions.
It is evident that $H\left(  t\right)  $ is not an Hermitian operator when
$\omega_{t}\notin%
\mathbb{R}
$ and/or $\alpha_{t}\neq\beta_{t}^{\ast}$. It becomes $\mathcal{PT}$-symmetric
when demanding $\omega_{t}$ to be an even function in $t$ or a generic
function of $it$, simultaneously with demanding $\alpha_{t},\beta_{t}$ to be
odd functions in $t$ or pure-imaginary generic functions of $it$.

Next, we consider the case where the TD Dyson map $\eta_{t}$ is not derived
from the Schr\"{o}dinger-like equation (\ref{eq5}), so that the Hamiltonian
$H_{t}$ is not an observable and a whole chain of gauge linked Hamiltonians
can be derived from it. Thus, considering the following ansatz for an
Hermitian time-dependent Dyson map
\begin{equation}
\bar{\eta}_{t}=\exp\left(  \bar{\gamma}_{t}a+\bar{\gamma}_{t}^{\ast}a^{\dag
}\right)  \text{,}\label{27}%
\end{equation}
we may construct the operator $\overset{\_}{H}_{t}$ using Eq. (\ref{eq17a}),
which will be fully defined only after the calculation of the time-dependent
parameters $\bar{\gamma}\left(  t\right)  $ and $\bar{\lambda}\left(
t\right)  $. Then, by transforming the Schr\"{o}dinger equation for
$\overset{\_}{H}_{t}$ through the Dyson map $\bar{\eta}_{t}$, we obtain the
Schr\"{o}dinger equation for
\begin{align}
\bar{h}_{t} &  =\bar{\eta}_{t}H_{t}\left(  \bar{\eta}_{t}\right)
^{-1}\nonumber\\
&  =\omega_{t}a^{\dagger}a+\bar{u}_{t}a+\bar{v}_{t}a^{\dagger}+\bar{f}%
_{t}\text{,}\label{28}%
\end{align}
with
\begin{subequations}
\begin{align}
\bar{u}_{t} &  =\omega_{t}\bar{\gamma}_{t}+\alpha_{t}\text{,}\label{29a}\\
\bar{v}_{t} &  =-\omega_{t}\left(  \bar{\gamma}_{t}\right)  ^{\ast}+\beta
_{t}\text{,}\label{29b}\\
\bar{f}_{t} &  =-\omega_{t}\left\vert \bar{\gamma}_{t}\right\vert ^{2}%
-\alpha_{t}\left(  \bar{\gamma}_{t}\right)  ^{\ast}+\beta_{t}\bar{\gamma}%
_{t}\text{.}\label{29c}%
\end{align}
By imposing $\bar{h}_{t}$ to be Hermitian, i.e., $\omega_{t},\bar{f}_{t}$ $\in%
\mathbb{R}
$ and $\bar{v}_{t}=\bar{u}_{t}^{\ast}$, we obtain $\bar{\gamma}_{t}=\left[
\beta_{t}^{\ast}-\alpha_{t}\right]  /2\omega_{t}$ and $\alpha_{t}\beta_{t}\in%
\mathbb{R}
$; consequently, it follows that $\bar{f}_{t}=\left[  \left\vert \alpha
_{t}\right\vert ^{2}+\left\vert \beta_{t}\right\vert ^{2}-2\alpha_{t}\beta
_{t}\right]  /4\omega_{t}$.

Next, we transform the Schr\"{o}dinger equation for $H_{t}$ using the
invariant form for the Hermitian time-dependent Dyson map%
\end{subequations}
\begin{equation}
\eta_{t}=\exp\left(  \gamma_{t}a+\gamma_{t}^{\ast}a^{\dag}\right)  \text{,}
\label{eq30}%
\end{equation}
to obtain the Schr\"{o}dinger equation for
\begin{align}
h_{t}  &  =\eta_{t}H_{t}^{\prime}\eta_{t}^{-1}\nonumber\\
&  =\omega_{t}a^{\dagger}a+u_{t}a+v_{t}a^{\dagger}+f_{t}\text{,} \label{eq31}%
\end{align}
with $H_{t}^{\prime}$ given by Eq. (\ref{9}) and
\begin{subequations}
\label{32}%
\begin{align}
u_{t}  &  =\omega_{t}\gamma_{t}+\alpha_{t}+i\partial_{t}\gamma_{t}%
\text{,}\label{32a}\\
v_{t}  &  =-\omega_{t}\gamma_{t}^{\ast}+\beta_{t}+i\partial_{t}\gamma
_{t}^{\ast}\text{,}\label{32b}\\
f_{t}  &  =-\omega_{t}\left\vert \gamma_{t}\right\vert ^{2}-\alpha_{t}%
\gamma_{t}^{\ast}+\beta_{t}\gamma_{t}+\frac{i}{2}\left(  \gamma_{t}%
\partial_{t}\gamma_{t}^{\ast}-\gamma_{t}^{\ast}\partial_{t}\gamma_{t}\right)
\text{.} \label{32c}%
\end{align}
For $h_{t}$ to be Hermitian we require $\bar{f}_{t}$ $\in%
\mathbb{R}
$ and $v_{t}=u_{t}^{\ast}$ what lead to the equation $\partial_{t}\gamma
_{t}=i\omega_{t}\gamma_{t}+i\left[  \alpha_{t}-\beta_{t}^{\ast}\right]  /2$,
and consequently to $u_{t}=\bar{u}_{t}=v_{t}^{\ast}=\bar{v}_{t}^{\ast}=\left[
\alpha_{t}+\beta_{t}^{\ast}\right]  /2$ and $f_{t}=-\alpha_{t}\gamma_{t}%
^{\ast}+\beta_{t}\gamma_{t}+\operatorname{Re}\left[  \alpha_{t}\gamma
_{t}^{\ast}-\beta_{t}\gamma_{t}\right]  /2$, such that%
\end{subequations}
\begin{equation}
h_{t}=\bar{h}_{t}+\mathcal{\bar{C}}_{t}\text{,} \label{34}%
\end{equation}
with $\mathcal{\bar{C}}_{t}=f_{t}-\bar{f}_{t}$.

We can go further by computing $h_{t}^{\prime}=\eta_{t}^{\prime}H_{t}%
^{\prime\prime}\left(  \eta_{t}^{\prime}\right)  ^{-1}$ from the invariant
form $\eta_{t}^{\prime}=\exp\left[  \gamma_{t}^{\prime}a+\left(  \gamma
_{t}^{\prime}\right)  ^{\ast}a^{\dag}\right]  $ and $H_{t}^{\prime\prime}$
given by Eq.(\ref{14a}); we obtain
\begin{equation}
h_{t}^{\prime}=\omega_{t}a^{\dagger}a+u_{t}a+v_{t}a^{\dagger}+f_{t}^{\prime
}\text{,}\label{35}%
\end{equation}
with
\begin{align}
f_{t}^{\prime} &  =\omega_{t}\left(  \left\vert \gamma_{t}\right\vert
^{2}+2\left\vert \gamma_{t}^{\prime}\right\vert ^{2}-\operatorname{Re}\left[
\gamma_{t}\left(  \gamma_{t}^{\prime}\right)  ^{\ast}\right]  \right)
\nonumber\\
&  +\frac{1}{2}\operatorname{Re}\left(  \alpha_{t}\gamma_{t}^{\ast}-\beta
_{t}\gamma_{t}\right)  -i\operatorname{Im}\left[  \alpha_{t}\left(  \gamma
_{t}^{\prime}\right)  ^{\ast}-\beta_{t}\gamma_{t}^{\prime}\right]
,\label{eq35}%
\end{align}
such that, similarly to Eq. (\ref{34}), it follows that
\begin{equation}
h_{t}^{\prime}=h_{t}+\mathcal{C}_{t}\text{,}\label{36}%
\end{equation}
with $\mathcal{C}_{t}=f_{t}^{\prime}-f_{t}$.

Therefore, the gauge operators $...,\bar{A}_{t},A_{t},...$ connecting the
adjacent spaces $...,\mathcal{H}(\bar{\phi}),\mathcal{H}(\phi),\mathcal{H}%
(\phi^{\prime}),...$ are global operators of the form $...,\bar{A}_{t}%
=\exp\left(  i%
{\textstyle\int\nolimits_{0}^{t}}
\mathcal{\bar{C}}_{\tau}d\tau\right)  ,A_{t}=\exp\left(  i%
{\textstyle\int\nolimits_{0}^{t}}
\mathcal{C}_{\tau}d\tau\right)  ,...$ which makes the matrix elements of the
observables $...,\bar{O}_{t}=\bar{\eta}_{t}^{-1}o_{t}\bar{\eta}_{t},O_{t}%
=\eta_{t}^{-1}o_{t}\eta_{t},...$ exactly the same in whatever the Hilbert
space $...\mathcal{H}(\bar{\phi}),\mathcal{H}(\phi),\mathcal{H}(\phi^{\prime
}),...$. Considering, for example, the field quadratures%
\begin{equation}
x_{k}=\frac{a-(-1)^{k}a^{\dagger}}{2i^{k-1}},\text{ \ \ }k=1,2,\label{37}%
\end{equation}
we obtain the observables
\begin{subequations}
\label{38}%
\begin{align}
\bar{X}_{k,t} &  =\left(  \bar{\eta}_{t}\right)  ^{-1}x_{k}\bar{\eta}%
_{t}=x_{k}+\frac{\bar{\gamma}_{t}+(-1)^{k}\left(  \bar{\gamma}_{t}\right)
^{\ast}}{2i^{k-1}}\text{,}\label{38a}\\
X_{k,t} &  =\eta_{t}^{-1}x_{k}\eta_{t}=x_{k}+\frac{\gamma_{t}+(-1)^{k}%
\gamma_{t}^{\ast}}{2i^{k-1}}\text{,}\label{38b}%
\end{align}
such that the quasi-Hermitian operators $\bar{X}_{1,t},X_{1,t}$ $\left(
\bar{X}_{2,t},X_{2,t}\right)  $ equal their $L^{2}$-counterparts except for
time-dependent functions which equal zero when $\bar{\gamma}_{t},\gamma_{t}$
go through a pure real (pure imaginary). To compute the matrix elements of
these observables we follow the reasonings in Ref. \cite{PL}, where the
solution of the Schr\"{o}dinger equation for the Hermitian TD Hamiltonian is a
displaced Fock state $\left\vert m\right\rangle $ apart from a TD global phase factor%

\end{subequations}
\begin{equation}
\left\vert \phi_{m,t}\right\rangle =e^{i\varphi_{m,t}}D\left(  \theta
_{t}\right)  \left\vert m\right\rangle \text{,} \label{39}%
\end{equation}
where $D\left(  \theta_{t}\right)  =\exp\left[  \theta_{t}a^{\dag}-\theta
_{t}^{\ast}a\right]  $ is the displacement operator, with $\theta_{t}%
=\theta_{0}\exp\left(  -i\chi_{t}\right)  $, $\chi_{t}=\int_{0}^{t}%
\omega_{\tau}d\tau$, and
\begin{align}
\varphi_{m,t}  &  =%
{\textstyle\int\nolimits_{0}^{t}}
d\tau\left\langle m\right\vert D^{\dagger}\left(  \theta_{\tau}\right)
\left(  i\partial_{\tau}-h_{\tau}\right)  D\left(  \theta_{\tau}\right)
\left\vert m\right\rangle \nonumber\\
&  =-m\chi_{t}-%
{\textstyle\int\nolimits_{0}^{t}}
d\tau f_{\tau}\text{.} \label{40}%
\end{align}
is the well-known Lewis and Riesenfeld phase \cite{LR}. The state vector
(\ref{39}) can be conveniently rewritten as%

\begin{equation}
\left\vert \phi_{m,t}\right\rangle =\Upsilon_{t}D\left(  \theta_{t}\right)
R\left(  \omega_{t}\right)  \left\vert m\right\rangle , \label{41}%
\end{equation}
where $\Upsilon_{t}=\exp\left(  -i%
{\textstyle\int\nolimits_{0}^{t}}
d\tau f_{\tau}\right)  $ is a global phase factor and $R\left(  \omega
_{t}\right)  =\exp\left(  -i\chi_{t}a^{\dagger}a\right)  $ a rotation
operator. Thus, for a generic superposition $\left\vert \phi_{t}\right\rangle
=%
{\textstyle\sum\nolimits_{m}}
c_{m}\left\vert \phi_{m,t}\right\rangle $ it follows that $\left\vert \phi
_{t}\right\rangle =U_{t}\left\vert \phi_{0}\right\rangle =\bar{A}%
_{t}\left\vert \bar{\phi}_{t}\right\rangle $, with the evolution operator%
\begin{equation}
U_{t}=\Upsilon_{t}D\left(  \theta_{t}\right)  R\left(  \omega_{t}\right)
D^{\dag}\left(  \theta_{0}\right)  \text{.} \label{42}%
\end{equation}

The base functions for all other Hermitian counterparts of the non-Hermitian
Hamiltonians, differ from those for $h_{t}$ ---as given by Eq. (\ref{41})---
only for the global phase factor $\Upsilon_{t}$, where the TD function $f_{t}$
must be replaced by the corresponding function related to the Hermitian
Hamiltonian. The TD frequency $\omega_{t}$ remains unchanged for all Hermitian
Hamiltonians as well as the TD parameter $\theta_{t}$.

For the initial coherent state $\left\vert \phi_{0}\right\rangle $, the
expectation values of the observables $\bar{O}_{t}$ and $O_{t}$ are thus
related with their Hermitian counterparts as%
\begin{equation}
\left\langle \bar{\psi}_{t}\right\vert \bar{O}_{t}\left\vert \bar{\psi}%
_{t}\right\rangle _{\bar{\rho}_{t}}=\left\langle \psi_{t}\right\vert
O_{t}\left\vert \psi_{t}\right\rangle _{\rho_{t}}=\left\langle e^{-i\chi_{t}%
}\left(  \phi_{0}-\theta_{0}\right)  +\theta_{t}\right\vert o_{t}\left\vert
e^{-i\chi_{t}}\left(  \phi_{0}-\theta_{0}\right)  +\theta_{t}\right\rangle
\text{,} \label{43}%
\end{equation}
leading to the expectation values of the quadratures
\begin{subequations}
\label{44}%
\begin{align}
\left\langle \bar{\psi}_{t}\right\vert \bar{X}_{1,t}\left\vert \bar{\psi}%
_{t}\right\rangle _{\bar{\rho}_{t}}  &  =\left\langle \psi_{t}\right\vert
X_{1,t}\left\vert \psi_{t}\right\rangle _{\bar{\rho}_{t}}=\operatorname{Re}%
\left\{  e^{-i\chi_{t}}\left(  \phi_{0}-\theta_{0}\right)  +\theta
_{t}\right\}  \text{,}\label{44a}\\
\left\langle \bar{\psi}_{t}\right\vert \bar{X}_{2,t}\left\vert \bar{\psi}%
_{t}\right\rangle _{\bar{\rho}_{t}}  &  =\left\langle \psi_{t}\right\vert
X_{2,t}\left\vert \psi_{t}\right\rangle _{\bar{\rho}_{t}}=\operatorname{Im}%
\left\{  e^{-i\chi_{t}}\left(  \phi_{0}-\theta_{0}\right)  +\theta
_{t}\right\}  \text{.} \label{44b}%
\end{align}

\subsection{Local gauges: The generalized time-dependent Swanson Hamiltonian}

We now address the case where a TD harmonic oscillator is under a TD
non-Hermitian parametric amplification process, which correspond to the
generalized time-dependent Swanson Hamiltonian \cite{Miled}%
\end{subequations}
\begin{equation}
H_{t}=\omega_{t}\left(  a^{\dagger}a+1/2\right)  +\alpha_{t}a^{2}+\beta
_{t}a^{\dagger2}\text{,}\label{45}%
\end{equation}
where $\omega_{t},\alpha_{t},\beta_{t}\in\mathbb{C}$. When $\omega_{t}%
\notin\mathbb{R}$ or $\alpha_{t}\neq\beta_{t}^{\ast}$ the above Hamiltonian is
clearly not Hermitian, and it becomes $\mathcal{PT}$-symmetric when demanding
$\omega_{t},\alpha_{t},\beta_{t}$ to be even functions in $t$ or generic
functions of $it$. Considering the Hermitian time-dependent Dyson map
\begin{align}
\bar{\eta}_{t} &  =\exp\left[  \bar{\epsilon}_{t}\left(  a^{\dagger
}a+1/2\right)  +\bar{\mu}_{t}a^{2}+\left(  \bar{\mu}_{t}\right)  ^{\ast
}a^{\dagger2}\right]  \nonumber\\
&  =\exp\left(  \bar{\lambda}_{+,t}K_{+}\right)  \exp\left(  \ln\bar{\lambda
}_{0,t}K_{0}\right)  \exp\left(  \bar{\lambda}_{-,t}K_{-}\right)
\text{.}\label{46}%
\end{align}
where $K_{+}=a^{\dagger2}/2$, $K_{-}=a^{2}/2$, $K_{0}=(a^{\dagger}a/2+1/4)$
form an $SU(1,1)$-algebra, with the TD coefficients
\begin{subequations}
\label{8}%
\begin{align}
\bar{\lambda}_{+,t} &  =-\bar{\Phi}_{t}e^{-i\bar{\varphi}_{t}}\text{,}%
\label{8a}\\
\bar{\lambda}_{-,t} &  =-\bar{\Phi}_{t}e^{i\bar{\varphi}_{t}}\text{,}%
\label{8b}\\
\bar{\lambda}_{0,t} &  =\bar{\Phi}_{t}^{2}-\bar{\chi}_{t}\text{.}\label{8c}%
\end{align}
where $\bar{\Phi}_{t}=\left\vert \bar{z}_{t}\right\vert /\bar{\Gamma}_{-,t}$
and $\bar{\chi}_{t}=2\bar{\Phi}_{t}/\left\vert \bar{z}_{t}\right\vert -1$,
with $\bar{\Gamma}_{\pm,t}=1\pm\left(  \bar{\Xi}_{t}/\bar{\epsilon}%
_{t}\right)  \coth\bar{\Xi}_{t}$, $\bar{\Xi}_{t}=\sqrt{\bar{\epsilon}_{t}%
^{2}-4\left\vert \bar{\mu}_{t}\right\vert ^{2}}$, and $\bar{z}_{t}=2\bar{\mu
}_{t}/\bar{\epsilon}_{t}=\left\vert \bar{z}_{t}\right\vert e^{i\bar{\varphi
}_{t}}$. Now, using the relation
\end{subequations}
\begin{equation}
\bar{\eta}_{t}%
\begin{pmatrix}
a\\
a^{\dagger}%
\end{pmatrix}
\bar{\eta}_{t}^{-1}=\pm\frac{1}{\sqrt{\bar{\lambda}_{0,t}}}%
\begin{pmatrix}
-1 & \bar{\lambda}_{+,t}\\
-\bar{\lambda}_{-,t} & \bar{\chi}_{t}%
\end{pmatrix}%
\begin{pmatrix}
a\\
a^{\dagger}%
\end{pmatrix}
\text{,}\label{47}%
\end{equation}
we obtain the transformed Hamiltonian
\begin{align}
\bar{h}_{t} &  =\bar{\eta}_{t}H_{t}\bar{\eta}_{t}^{-1}\nonumber\\
&  =\bar{W}_{t}(a^{\dagger}a+1/2)+\bar{V}_{t}a^{2}+\bar{T}_{t}a^{\dagger
2}\text{.}\label{10}%
\end{align}
where, by defining $\omega_{t}=\left\vert \omega_{t}\right\vert e^{i\varphi
_{\omega,t}}$, $\alpha_{t}=\left\vert \alpha_{t}\right\vert e^{i\varphi
_{\alpha,t}}$, and $\beta_{t}=\left\vert \beta_{t}\right\vert e^{i\varphi
_{\beta,t}}$, the coefficient functions which assure the Hermiticity of
$\bar{h}_{t}$, i.e., $\bar{W}_{t}\in%
\mathbb{R}
$ and $\bar{T}_{t}=\bar{V}_{t}^{\ast}$, are given by
\begin{subequations}
\label{48}%
\begin{align}
\bar{W}_{t} &  =\frac{1}{\bar{\chi}_{t}-\bar{\Phi}_{t}^{2}}\left\{  \left\vert
\omega_{t}\right\vert \left(  \bar{\chi}_{t}+\bar{\Phi}_{t}^{2}\right)
\cos\varphi_{\omega,t}-2\bar{\Phi}_{t}\left[  \left\vert \alpha_{t}\right\vert
\cos\left(  \bar{\varphi}_{t}-\varphi_{\alpha,t}\right)  +\left\vert \beta
_{t}\right\vert \bar{\chi}_{t}\cos\left(  \bar{\varphi}_{t}+\varphi_{\beta
,t}\right)  \right]  \right\}  \text{,}\label{48a}\\
\bar{V}_{t} &  =\frac{1}{\bar{\chi}_{t}-\bar{\Phi}_{t}^{2}}\left(  \left\vert
\omega_{t}\right\vert \bar{\Phi}_{t}e^{i\left(  \bar{\varphi}_{t}%
+\varphi_{\omega,t}\right)  }-\left\vert \alpha_{t}\right\vert e^{i\varphi
_{\alpha,t}}-\left\vert \beta_{t}\right\vert \bar{\Phi}_{t}^{2}e^{-2i\bar
{\varphi}_{t}}\right)  \text{,}\label{48b}%
\end{align}
with $\bar{\Phi}_{t}$ and $\bar{\varphi}_{t}$ following from the system
\end{subequations}
\begin{subequations}
\label{49}%
\begin{align}
\left[  \left\vert \omega_{t}\right\vert \bar{\Phi}_{t}\sin\varphi_{\omega
,t}+\left\vert \alpha_{t}\right\vert \sin\left(  \bar{\varphi}_{t}%
-\varphi_{\alpha,t}\right)  \right]  \left(  1-\bar{\Phi}_{t}^{2}\right)
+\left\vert \beta_{t}\right\vert \left[  \left(  2\bar{\chi}_{t}-1\right)
\bar{\Phi}_{t}^{2}-\bar{\chi}_{t}^{2}\right]  \sin\left(  \bar{\varphi}%
_{t}+\varphi_{\beta,t}\right)   &  =0\text{,}\label{49a}\\
\left(  \bar{\chi}_{t}-1\right)  \bar{\Phi}_{t}\left\vert \omega
_{t}\right\vert \cos\varphi_{\omega,t}+\left\vert \alpha_{t}\right\vert
\left(  1-\bar{\Phi}_{t}^{2}\right)  \cos\left(  \bar{\varphi}_{t}%
-\varphi_{\alpha,t}\right)  +\left\vert \beta_{t}\right\vert \left(  \bar
{\Phi}_{t}^{2}-\bar{\chi}_{t}^{2}\right)  \cos\left(  \bar{\varphi}%
_{t}+\varphi_{\beta,t}\right)   &  =0\text{.}\label{49b}%
\end{align}

Next, under the invariant Dyson map
\end{subequations}
\begin{equation}
\eta_{t}=\exp\left(  \lambda_{+,t}K_{+}\right)  \exp\left(  \ln\lambda
_{0,t}K_{0}\right)  \exp\left(  \lambda_{-,t}K_{-}\right)  \text{,}\label{50}%
\end{equation}
we derive the Hamiltonian
\begin{align}
h_{t} &  =\eta_{t}H_{t}\eta_{t}^{-1}+i\dot{\eta}_{t}\eta_{t}^{-1}\nonumber\\
&  =W_{t}(a^{\dagger}a+1/2)+V_{t}a^{2}+T_{t}a^{\dagger2}\text{,}\label{51}%
\end{align}
where the coefficient functions which assure its Hermiticity are given by
\begin{subequations}
\label{52}%
\begin{align}
W_{t} &  =\left\vert \omega_{t}\right\vert \cos\varphi_{\omega,t}+\frac
{2\Phi_{t}}{1-\chi_{t}}\left[  \left\vert \alpha_{t}\right\vert \cos\left(
\varphi_{t}-\varphi_{\alpha,t}\right)  -\left\vert \beta_{t}\right\vert
\cos\left(  \varphi_{t}+\varphi_{\beta,t}\right)  \right]  \text{,}%
\label{52a}\\
V_{t} &  =\frac{1}{1-\chi_{t}}\left(  \left\vert \alpha_{t}\right\vert
e^{i\varphi_{\alpha,t}}-\left\vert \beta_{t}\right\vert \chi_{t}%
e^{-i\varphi_{\beta,t}}-i\left\vert \omega_{t}\right\vert \Phi_{t}\sin
\varphi_{\omega,t}e^{i\varphi t}\right)  \text{.}\label{52b}%
\end{align}
with $\Phi_{t}$ and $\varphi_{t}$ following from the couple nonlinear
equations
\end{subequations}
\begin{subequations}
\label{54}%
\begin{align}
\dot{\Phi}_{t} &  =\frac{2}{\chi_{t}-1}\left\{  \left[  \left\vert \omega
_{t}\right\vert \Phi_{t}\sin\varphi_{\omega,t}+\left\vert \alpha
_{t}\right\vert \sin\left(  \varphi_{t}-\varphi_{\alpha,t}\right)  \right]
\left(  1-\Phi_{t}^{2}\right)  +\right.  \nonumber\\
&  \left.  \left\vert \beta_{t}\right\vert \left[  \left(  2\chi_{t}-1\right)
\Phi_{t}^{2}-\chi_{t}^{2}\right]  \sin\left(  \varphi_{t}+\varphi_{\beta
,t}\right)  \right\}  \label{54a}\\
\dot{\varphi}_{t} &  =\frac{2}{\left(  \chi_{t}-1\right)  \Phi_{t}}\left[
\left\vert \alpha_{t}\right\vert \left(  1-\Phi_{t}^{2}\right)  \cos\left(
\varphi_{t}-\varphi_{\alpha,t}\right)  \right.  \left.  +\left\vert \beta
_{t}\right\vert \left(  \Phi_{t}^{2}-\chi_{t}^{2}\right)  \cos\left(
\varphi_{t}+\varphi_{\beta,t}\right)  \right]  \nonumber\\
&  +2\left\vert \omega_{t}\right\vert \cos\varphi_{\omega,t}\text{,}%
\label{54b}%
\end{align}
which automatically reduces to those in Eq. (\ref{49}) when considering
$\dot{\Phi}_{t}=\dot{\varphi}_{t}=0$. It is immediate to conclude that in this
case the difference $h_{t}-$ $\bar{h}_{t}$ is not a c-number, and
consequently, a local gauge transformation is required to link these two
Hermitian Hamiltonians.

\section{Conclusions}

We have here considered the problem of TD non-Hermitian Hamiltonians under TD
Dyson maps, a subject of significant importance that has received attention in
the recent literature on non-Hermitian quantum mechanics
\cite{Mostafa,Fring,Miled,Luiz}. We have first shown how to construct from a
given TD non-observable non-Hermitian Hamiltonian an infinite chain of gauge
linked Hamiltonians, whose associated observables and their matrix elements
are all related to each other. In the particular case where the Hamiltonians
are linked together by global gauges, the matrix elements of the observables
associated with these TD non-observables non-Hermitian Hamiltonians became all
identical to each other, making all these Hamiltonians equivalents. In such a
case, by approaching whatever the Hamiltonian in the chain we can get
information about any other Hamiltonian, and this property becomes all the
more important when we find that among the infinite Hamiltonians in the chain
one of them is definitely easier to treat: $\bar{H}_{t}$, whose Hermitian
counterpart, $\bar{h}_{t}=\bar{\eta}_{t}H_{t}\left(  \bar{\eta}_{t}\right)
^{-1}$, does not demand the time derivative of the corresponding Dyson map,
$\bar{\eta}_{t}$, and consequently, the Gauss decomposition of this operator.

When, on the other hand, we ensure these TD non-Hermitian Hamiltonian to be
observables, simultaneously to ensuring the unitarity of the time-evolution
they govern, the whole chain collapse to a single TD observable non-Hermitian
Hamiltonian. Therefore, the observability character of a TD non-Hermitian
Hamiltonian prevents the possibility of gauge-linked associated Hamiltonians
and observables. After going through these properties, we then present two
illustrative examples: the TD harmonic oscillators under TD linear and
parametric non-Hermitian amplification processes, the linear case resulting in
global gauge transformations.

The properties derived here help us to better understand systems described by
time-dependent non-Hermitian Hamiltonians under time-dependent Dyson maps,
which have only recently been studied and should play a central role in
non-Hermitian quantum mechanics.
\end{subequations}
\begin{flushleft}
{\Large \textbf{Acknowledgements}}
\end{flushleft}

FSL would like to thank CNPq (Brazil) for support, and MHYM would like to
thank CAPES (Brazil) for support and the City University London for kind
hospitality. The authors are in debt to Prof. A. Fring of the City University
London, for many fruitful discussions.

\end{document}